\documentclass[10pt]{article}

\usepackage{amsmath}

\usepackage{array}

\usepackage{appendix}

\usepackage{tocloft}                   

\usepackage{graphicx}

\usepackage{amsfonts}

\usepackage{amssymb}

\usepackage{mathrsfs}

\usepackage{yfonts}

\usepackage{euscript}

\usepackage{centernot}                 

\usepackage{ifsym}                     

\usepackage{upgreek}

\usepackage{mathtools}

\usepackage{color}

\usepackage{slantsc}
\usepackage{calligra}

\usepackage{bbold}          

\usepackage[T1]{fontenc}

\usepackage{epsf}

\usepackage{latexsym}

\usepackage{tipa}

\usepackage{makeidx}

\makeindex

\def\b{\begin{equation}}

\def\e{\begin{equation}}

\def\be{\begin{equation}}              

\def\ee{\end{equation}}

\def\beq{\begin{equation}}

\def\eeq{\end{equation}}

\def\bea{\begin{eqnarray}}

\def\eea{\end{eqnarray}}

\def\m{\mbox{ }}

\def\mma {\m , \m \m }

\def\!{\hspace{-1.6667em}}

\def\n{\noindent}

\def\u{\underline}

\def\disjoint{\mbox{\scriptsize $\coprod$}}

\def\me{\mbox{e}}

\def\mg{\mbox{g}}

\def\mh{\mbox{h}}

\def\ml{\mbox{l}}

\def\mp{\mbox{p}}

\def\bh{\u{\u{\mbox{h}}}  }            

\def\bA{\mbox{\bf A}}

\def\bC{\mbox{\bf C}}                    

\def\bE{\mbox{\bf E}}

\def\bG{\mbox{\bf G}}                     

\def\bP{\mbox{\bf P}}

\def\bS{\mbox{\bf S}}

\def\bg{\mbox{\bf g}}

\def\bh{\mbox{\bf h}}

\def\buu{\mbox{\bf u}}             

\def\bupSigma{\mbox{\boldmath$\Sigma$}}                 

\def\fC{\mbox{\sffamily C}}

\def\fD{\mbox{\sffamily D}}

\def\cK{{\mathscr K}}

\def\cL{{\mathscr L}}

\def\sp{\mbox{\scriptsize p}}

\def\sq{\mbox{\scriptsize q}}

\def\sF{\mbox{\scriptsize F}}

\def\sumi2{\sum\mbox{}_{\mbox{}_{\mbox{\scriptsize $i$=1}}}^2}

\def\sumi3{\sum\mbox{}_{\mbox{}_{\mbox{\scriptsize $i$=1}}}^3}

\def\sumABcycles3{\sum\mbox{}_{\mbox{}_{\mbox{\scriptsize cycles $A,B$=1}}}^{3}}

\def\sumCDcycles3{\sum\mbox{}_{\mbox{}_{\mbox{\scriptsize cycles $C,D$=1}}}^{3}}

\def\sumj3{\sum\mbox{}_{\mbox{}_{\mbox{\scriptsize $j$=1}}}^3}

\def\sumk3{\sum\mbox{}_{\mbox{}_{\mbox{\scriptsize $k$=1}}}^3}

\def\prodiA1{\prod\mbox{}_{\mbox{}_{\mbox{\scriptsize $i$=1}}}^{A - 1}}

\def\bigtimes{\mbox{\Large $\times$}}

\def\es{\m = \m}

\def\:={\m := \m}

\def\=:{\m =: \m}

\def\FrC{\mbox{$\mathfrak{C}$}}                                
                                                       
\def\FrX{\mathfrak{X}}                                         
\def\FrS{\mbox{\Large $\mathfrak{s}$}}                         
\def\FrU{\mbox{$\mathfrak{U}$}}                                
\def\FrV{\mbox{$\mathfrak{V}$}}                                
\def\FrW{\mbox{$\mathfrak{W}$}}                                
\def\FrM{\mbox{$\mathfrak{M}$}}                                
\def\FrN{\mbox{$\mathfrak{N}$}}                                
\def\lFrg{\mbox{\Large$\mathfrak{g}$}}                         
\def\FrN{\mathfrak{N}}                                         
                                                  
\def\FrG{\mathfrak{G}}                                         
\def\Hilb{\mbox{{\boldmath$\mathfrak{H}$}ilb}}                 
\def\FrQ{\mbox{\Large $\mathfrak{q}$}}                               
\def\bFrC{\mbox{\boldmath$\mathfrak{C}$}}                            
\def\Phase{\mbox{{\boldmath$\mathfrak{P}$}hase}}                     

\def\bFrR{\mbox{\boldmath$\mathfrak{R}$}}                            
\def\Rig-Phase{\bFrR\mbox{ig-}\Phase}                                
                                                                													   
\def\FrP{\mbox{\Large $\mathfrak{p}$}}                                 
\def\FrR{\mbox{\boldmath$\mathfrak{R}$}}                             

\def\bFrR{\mbox{\boldmath$\mathfrak{R}$}}                            

\def\bFrR{\mbox{\boldmath$\mathfrak{R}$}}                            

\def\1mat{\u{\u{1}}}                                                 

\def\Positive-Modespace{\mbox{{\boldmath$\mathfrak{M}$}odespace$^+$}}

\def\POSITIVE-MODESPACE{\mbox{{\boldmath$\mathfrak{M}$}ODESPACE$^+$}}
                                                                                                                             														
\def\bFrS{\mbox{\Large $\mathfrak{s}$}}                              
			
\def\blFrS{\mbox{\LARGE $\mathfrak{s}$}}                             
\def\Riem{\bFrR\mbox{iem}}                                           
\def\CRiem{\bFrC\Riem}                                               

\def\Superspace{\bFrS\mbox{uperspace}}                               
\def\lSuperspace{\blFrS\mbox{uperspace}}                             

\def\CS{\bFrC\bFrS}                                                  

\def\Rel{\FrR\me\ml}

\def\lattice{\mbox{\bf\Large$\mathfrak{L}$}}                                      

\def\Kin-Hilb{\mbox{{\boldmath$\mathfrak{K}$}in-\Hilb}}                     

\def\Mid-Hilb{\mbox{{\boldmath$\mathfrak{M}$}id-\Hilb}}                     

\def\Dyn-Hilb{\mbox{{\boldmath$\mathfrak{D}$}yn-\Hilb}}                     

\def\5Star{\mbox{\Large$\star$}}              

\begin{document}

\begin{titlepage}

\begin{center}

{\bf\Large Absolute versus Relational Debate:} 

\vspace{0.1in}

{\bf\Large a Modern Global Version}

\m

{\bf Dr Edward Anderson}$^*$  

\m 

\end{center}

\begin{abstract}

\n Suppose one seeks to free oneself from a symmetric absolute space by quotienting out its symmetry group. 
This in general however fails to erase all memory of this absolute space's symmetry properties.  
Stratification is one major reason for this, which is present in both a) Kendall-type Shape Theory and subsequent Relational Mechanics, 
                                                                  and b) General Relativity configuration spaces.  
We consider the alternative starting point with a generic absolute space, meaning with no nontrivial generalized Killing vectors whatsoever. 
In this approach, generically Shape-and-Scale Theory is but trivially realized, there is no separate Shape Theory and indeed no stratification.  
While the GR configuration space version of these considerations was already expounded in 1996 by Fischer and Moncrief, 
the Kendall-type shape theory version is new to the current article.
In each case, this amounts to admitting some small deformation by which symmetry's hard consequences at the level of reduced configuration spaces are warded off.
We end by discussing the senses in which each of the above two strategies retain absolutist features, 
each's main known technical advantages and disadvantages, and the desirability of replacing Kendall-type Shape Theory with a Local-and-Approximate Shape Theory.  
This article is in honour of Prof. Niall \'{o} Murchadha, on the occasion of his Festschrift.

\end{abstract}

\m 

\n PACS: 04.20.Cv, 02.40.-k 

\m 

\n Physics keywords: Configuration spaces, GR superspace and conformal superspace, $N$-body problem, Relationalism, Background Independence.

\m

\n Mathematics keywords: Generalized Killing equations, stratified manifolds, Hausdorff versus Kolmogorov notions of separability, Shape Theory, Shape Statistics.

\m

\n $^*$ dr.e.anderson.maths.physics *at* protonmail.com . Work done while at DAMTP, Cambridge.  

\end{titlepage}

\section{Introduction}

Suppose one seeks to free oneself from a symmetric absolute space by quotienting out its automorphism group.
This addresses the relational horn  \cite{BB82, B03, FileR, AMech, ABook} of the Absolute versus Relational (Motion) Debate \cite{Newton, L, M, DoD-Buckets}, 
often contemporarily also referred to as `seeking Background Independence' \cite{A64-A67, Giu06, ABook}.     
This seeking can be either indirect by applying a `Best Matching' \cite{BB82, B03} group action on unreduced configuration spaces 
or by direct formulation on reduced configuration spaces \cite{FORD, FileR}: `relational spaces'.  
Kendall's Shape Theory \cite{Kendall84, Kendall89, Kendall} (see also \cite{Small, JM00, Bhatta, DM16, PE16} for reviews)
is a trove of reduced configuration space geometry work greatly advancing Relational Mechanics 
(see \cite{FileR, ABook, A-Series, A-Monopoles}; \cite{Smale70, LR95-97, M02-M05} are related works). 
Both approaches are additionally amenable to removal of absolute time as well -- Temporal Relationalism -- \cite{BB82, B94I, ABook}; 
there is also a Spacetime Relationalism \cite{APoT3, ABook, Project-1} `freeing from absolute spacetime' counterpart.  

\m 

\n There is moreover a formulation of GR-as-geometrodynamics along such lines. 
\cite{BSW, B94I, BFO} provided some elements of this, 
from which \cite{ABFO} developed a formulation \cite{FileR, ABook} in which Temporal and Configurational Relationalism are implemented {\sl compatibly}. 
This is the first of many moves for combining individual resolutions of Problem of Time Facets \cite{K92-I93, APoT, APoT3} into a collective whole \cite{ABook}.
To make sense of this statement, 
let us point out that Configurational Relationalism \cite{ABook} itself being a substantial generalization of the `Thin Sandwich' \cite{BSW, BO, BF} Problem of Time facet. 
The automorphism group in this context is the spatial 3-diffeomorphisms, and the reduced space is Wheeler's superspace \cite{Battelle, DeWitt67, Fischer70, FM96, Giu09}.

\m  

\n In the context of the Initial Value Problem of GR \cite{York, OY} 
-- the principal research topic in the career of Niall \'{o} Murchadha, to whom this article is dedicated on the occasion of his Festschrift -- 
the spatial conformal transformations are furthermore quotiented out.
Aside from working on two of the previous paragraph's cited collaborations, Niall, Barbour, Foster, Kelleher and I gave a further relational formulation 
of GR-as-conformogeometrodynamics along such lines in \cite{ABFKO}.
The reduced space in this context is York's conformal superspace \cite{York}.  

\m

\n Kendall's Shape Theory itself concerns the similarity Killing equation, 
whereas Generalized-Kendall type Relational Theory involves the generalized Killing equation along the lines of Yano's reviews \cite{Yano} and outlined in Sec 2.    
(In the context of finite point-or-particle theories, we use `relational' to encompass both `shape' and `shape-and-scale'.)
In the GR case (Sec 4), the Killing equation plays both spacetime \cite{MacCallum} and geometrodynamical-space \cite{BO, Fischer70} roles, 
whereas the conformal Killing equation plays a significant role \cite{York, York74} in the GR Initial Value Problem.
On the other hand, affine and projective groups, and hence underlying affine and projective Killing equations are involved 
in some further generalized Kendall-type Relational Theories of particular note due to their foundational role in Image Analysis and Computer Vision \cite{Bhatta, PE16}.  

\m  

\n The current article concerns how such quotienting by automorphism groups however in general fails to erase all memory of this absolute space's symmetry properties.  
One reason for this follows from the (generalized) Killing equation's global sensitivity to the topology of the manifold it is acting upon.
A second reason concerns the stratification \cite{W-T, Whitney65, Thom69, 2000-Proceedings, Pflaum, Kreck} incurred by the quotienting; 
this occurs in both Kendall-type \cite{Kendall, GT09, KKH16} and GR \cite{Fischer70, Fischer86, FM96} contexts. 
In the recent article \cite{ACirc}, I isolated the first of these effects by comparing two absolute space models model exhibiting no strata: the circle versus the line. 
The current article complements this by considering stratificational reasons instead. 
On the one hand, Superspace and conformal superspace are stratified.
On the other hand, stratification can occur in similarity, Euclidean, affine and projective relational spaces, with varying severity, as laid out in Sec \ref{Strata}.  
This presentation is rooted in the theory of general topological {\sl spaces}, 
with the severity in question qualifying various stratified manifolds' departures from the physicists' comfort zone of topological {\sl manifolds}.  

\m 

\n The GR literature has moreover been pointing to an alternative perspective 
since Fischer and Moncrief's 1996 landmark treatment \cite{FM96} of superspace and conformal superspace.
Namely, that {\sl generic} topological manifolds' reduced configuration spaces exhibit no stratification. 
The corresponding point not having yet been formulated in Generalized Kendall-type Relational Theory or the associated field of Shape Statistics, 
the current article performs this function (Sec 5).
Thereby, finite point-or-particle models of Relationalism are vastly simpler for generic absolute space than for $\mathbb{R}^n$.
This is moreover a timely point due to considerable recent and ongoing expansion in the scope of Generalized Kendall-type Relational Theory 
\cite{Sparr, GT09, MP05, FileR, Bhatta, DM16, KKH16, PE16, ABook, ACirc, Top-Shapes, Project-1, Forth}.  
Stratification is a persistent consequence of symmetry, but can be avoided by building in a small defect in the first place, 
much as mediaeval cathedral builders incorporated a small `countertwist' so as to `ward off the devil'.  
The current article furthermore serves to point out that a wealth of Applied Geometry techniques habitually used in the study of GR 
also occupy a foundational role in the field of Shape Statistics \cite{Small, Kendall, JM00, Bhatta, DM16, PE16}.  
We also point to need for a local approximate theory of shapes, as outlined in the Conclusion (Sec 6). 
These are further Applied Geometry topics to which various past GR literature insights 
-- modelling approximate Killing vectors \cite{Matzner, AM78, C79, CKBook, Z-KN, H08} -- are welcomely applicable.

\vspace{10in}

\section{Finite case}\label{Finite}

\subsection{Absolute space and constellation space}

\n{\bf Definition 1} {\it Carrier space} $\FrC^d$, alias {\it absolute space} in the physically realized case, is an at-least-provisional model for the structure of space.

\m

\n{\bf Remark 1} While Geometry was originally conceived of as occurring in physical space or objects embedded therein (parchments, the surface of the Earth...), 
we consider the Geometry version of our problem in terms of the abstract carrier space rather than according it an absolute space interpretation.   
$\FrC^d$ can also be interpreted as a {\it sample space} in the context of Probability and Statistics, of {\it location data}.  

\m 

\n{\bf Remark 2} In some physical applications, the points model material particles (classical, and taken to be of negligible extent).
Because of this, we subsequently refer to constellations as consisting of points-or-particles. 

\m 

\n{\bf Definition 2} {\it Constellation space} is the product space 
\be 
\FrQ(\FrC^d, N)  \es  \bigtimes_{i = 1}^N \FrC^d                \m ,    
\ee 
where $N$ is the number of points-or-particles under consideration.

\subsection{Automorphism groups with $\mathbb{R}^d$-geometry examples}

\n{\bf Structure 1} Relational Theory furthermore considers some group of automorphisms 
\be
\lFrg = Aut(\langle \FrC^d, \sigma \rangle)
\ee 
of $\FrC^d$ [or $\FrQ(\FrC^d, N)$ by its product group structure] as equipped by some level of structure $\sigma$ to be irrelevant to the modelling in question. 
$\sigma$ is here some level of mathematical structure on $\FrC^d$.  

\m

\n{\bf Example 0} For $\sigma = \bh$ the metric structure on a manifold, 
\be 
Aut(\langle \FrC^d, \bh \rangle) = Isom(\FrC^d) \m : 
\ee 
the {\it isometry group} of $\FrC^d$. 
In the case of $\FrC^d = \mathbb{R}^d$, this is the {\it Euclidean group}   
\be 
Eucl(d) = Tr(d) \rtimes Rot(d)    
\ee 
of translations $Tr(d)$ and rotations $Rot(d)$, where $\rtimes$ denotes semidirect product of groups \cite{Cohn}.
This is the group most usually considered in the Absolute versus Relational Debate.  

\m 

\n{\bf Example 1} For $\sigma = \bS$ the metric structure up to constant rescalings, 
\be 
Aut(\langle \FrC^d, \bS \rangle) = Sim(\FrC^d) \m : 
\ee 
the {\it similarity group} of $\FrC^d$. 
In the case of $\FrC^d = \mathbb{R}^d$, this is the similarity group 
\be 
Sim(d) = Tr(d) \rtimes (Rot(d) \times Dil) \m 
\ee 
of translations, rotations and dilations $Dil$, where now $\times$ is the direct product of groups.  
This is the case standardly considered in Kendall's Shape Statistics \cite{Kendall84, Kendall89, Kendall}. 
It turns out moreover to be a useful intermediary and/or structure within Example 1's Mechanics context as well, as explained in Sec \ref{2.last}.

\m 

\n The invariants corresponding to this pair of theories are, respectively, inner products of differences in position coordinates and ratios thereof.

\m

\n{\bf Example 2} For $\sigma = \bA$ the affine structure, 
\be 
Aut(\langle \FrC^d, \bA \rangle) = Aff(\FrC^d) \m : 
\ee 
the {\it affine group} of $\FrC^d$. 
In the case of $\FrC^d = \mathbb{R}^d$, this is the affine group 
\be 
Aff(d) = Tr(d) \rtimes GL(d, \mathbb{R}) \m , 
\ee 
for $GL(d, \mathbb{R})$ the general-linear group of dilations, rotations, shears $Sh(d)$ and Procrustes stretches $Pr(d)$ \cite{Coxeter}. 
This is the case corresponding to Image Analysis from the idealized perspective of an infinitely distant observer \cite{Sparr, PE16}.  

\m 

\n{\bf Example 3} For $\sigma = \bE$ the equi-top-form structure, 
\be 
Aut(\langle \FrC^d, \bE \rangle) = Equi(\FrC^d) \m : 
\ee 
the {\it equi-top-form group} of $\FrC^d$. 
In the case of $\FrC^d = \mathbb{R}^d$, this is the equi-top-form group 
\be 
Equi(d) = Tr(d) \rtimes SL(d, \mathbb{R}) \m , 
\ee 
for $SL(d, \mathbb{R})$ the special-linear group of $Rot(d)$, $Sh(d)$ and $Pr(d)$. 
While $Eucl(d)$ corresponds to `interior alias dot product of difference vector' invariants and $Sim(d)$ to ratios thereof, 
$Equi(d)$ corresponds to `exterior product of difference vector' invariants and $Aff(d)$ to ratios thereof. 
(The 2-$d$ and 3-$d$ exterior products are the familiar cross product and scalar triple product; 
in 2-$d$, `equi-top-form' is `equiareal': a somewhat familiar type of geometry \cite{Coxeter}.)
In this sense $Equi(d)$ is to $Aff(d)$ exactly what $Eucl(d)$ is to $Sim(d)$.  

\m 

\n{\bf Example 4} For $\sigma = \bC$ the conformal structure, 
\be 
Aut(\langle \FrC^d, \bC \rangle) = Conf(\FrC^d) \m : 
\ee 
the {\it conformal group} of $\FrC^d$. 
In the case of $\FrC^d = \mathbb{R}^d$ for $d \geq 3$, this is the conformal group 
\be 
Conf(d) = SO(d, 1)
\ee
of $Tr(d)$, $Rot(d)$, $Dil$, and special conformal transformations $K(d)$ \cite{O17}. 
The invariants in this case are local angles.  

\m 

\n{\bf Remark 3} The above groups can be considered to be flat-geometrical knowns. 

\m 

\n{\bf Remark 4} We can moreover also {\sl solve} for such groups, as follows.

\subsection{Automorphism groups from solving generalized Killing equations}

\n{\bf Remark 5} For isometries on the differentiable manifold $\FrM^d$, 
the answer to this problem is very well-known in the GR community \cite{MacCallum}: one solves the Killing equation on $\FrM^d$,  
\be
\cK \, \xi = 0 \m , 
\ee
to obtain the Killing vectors $\xi$, and one then figures out from the Lie brackets between these which Lie group they form.   
In the case in hand, the model for absolute space $\FrC^d$ plays the part of $\FrM^d$.  

\m 

\n{\bf Remark 6} This line of reasoning moreover generalizes over 
all well-defined types of geometry arrived at from the preservation of levels of structure in excess of differentiable structure, as follows.  

\m 

\n We extend from the Killing equation -- corresponding to metric geometry -- to whichever generalized Killing equation \cite{Yano}, 
corresponding to whichever other level of geometrical structure. 
We proceed by concrete examples, then passing on to generalized observations. 
What we call $\cK$ above really means $\cK(\bh)$.  

\m 

\n{\bf Example 1} The {\it similarity Killing equation} on $\FrC^d$
\be
\cK(\bS) \, \xi = 0 
\ee 
is solved by similarity Killing vectors (on some occasions termed homothetic Killing vectors) forming the similarity group $Sim(\FrC^d)$.  

\m 

\n{\bf Example 2} The {\it affine Killing equation} on $\FrC^d$
\be 
\cK(\bA) \, \xi = 0
\ee 
is solved by affine Killing vectors, forming the affine group $Aff(\FrC^d)$.  

\m 

\n{\bf Example 3} The {\it equi-top-form Killing equation} on $\FrC^d$ 
\be 
\cK(\bE) \, \xi = 0
\ee 
is solved by equi-top-form Killing vectors, forming the affine group $Equi(\FrC^d)$. 

\m

\n{\bf Example 4} The {\it conformal Killing equation} on $\FrC^d$
\be 
\cK(\bC) \, \xi = 0 
\ee 
is solved by conformal Killing vectors, forming the conformal group $Conf(\FrC^d)$. 
(The differential operator $\cK(C)$ here is alias $\cL$ in the GR literature.)  

\m 

\n{\bf Example 5} The {\it projective Killing equation} on $\FrC^d$, for $\sigma = \bP$ the projective structure,
\be
\cK(\bP) \, \xi = 0
\ee 
is solved by projective Killing vectors, forming the projective group $Proj(\FrC^d)$. 

\m 

\n{\bf Remark 7} Specific forms for (almost all) of these equations can be found in \cite{Yano}. 
Their most notable feature is that they are all succinctly and intuitively expressible in terms of Lie derivatives.  
We write 
\be 
\cK(\sigma) \, \xi = 0 
\ee 
for the general meaningful notion of geometry on a differentiable manifold's {\it generalized Killing equation}.   
This notation moreover suppresses reference to the manifold $\FrM^d$; a full notation would be $\cK(\FrM^d, \sigma)$.
Due to the competing lattices of geometrical structures on the arbitrary differentiable manifold having the Lie derivative operation $\pounds_{\xi}$ available.
Furthermore, due to compatibility with the underlying differentiable manifold level, the generalized Killing equation is of the form 
\be 
\pounds_{\xi} \bG  =  f \, \bG  \m . 
\ee 
$\bG$ is here some geometrical object and $f$ is a scalar function.  
In some cases, the function is constant or zero.  
On the one hand, the metric, similarity and conformal classes `have metrics in the role of geometrical objects'. 
One can moreover rebrand geometrical objects to absorb such functions, 
for instance by considering similarity classes of metrics as geometrical objects rather than metrics themselves. 
On the other hand, affine, equiareal and projective classes `have connections in the role of geometrical objects'.

\subsection{Automorphism groups: topologically-identified and curved examples}

\n{\bf Example A} $\FrC^d = \mathbb{S}^1$ supports an isometry group
\be 
Isom(\mathbb{S}^1) = SO(2) 
                   = U(1) 
				   = \mathbb{S}^1                                                              \m , 
\ee 
where the last equality is at the level of manifolds. 
$\FrC^d = \mathbb{S}^1$ does not however support a distinct similarity group does not, since the generator of dilations does not respect the `periodic identification' of the circle.

\m 

\n{\bf Remark 8} $\mathbb{S}^1$ is moreover the first torus $\mathbb{T}^d$ and real-projective space $\mathbb{RP}^d$ as well as the first sphere $\mathbb{S}^d$; 
in Relational Theory, it turns out to behave most like the other $\mathbb{T}^d$.  

\m 

\n{\bf Example B} The flat $d$-dimensional torus $\mathbb{T}^d$ also supports an isometry group, 
\be 
Isom(\mathbb{T}^d) \es  \bigtimes_{\alpha = 1}^d U(1) 
                   \es  \bigtimes_{\alpha = 1}^d \mathbb{S}^1 
				   \es  \mathbb{T}^d                             \m ,
\ee 
but not a distinct similarity group, due likewise to incompatibility between the dilation generator and the underlying topological identification.

\m 

\n{\bf Example C} The $d$-dimensional sphere $\mathbb{S}^d$  $d \geq 2$ again supports an isometry group, 
\be 
Isom(\mathbb{S}^d) = SO(d + 1)                                                                                \m , 
\ee 
but not a distinct similarity group, this time due to incompatibility between the dilation generator and the sphere's curvature length scale.  
This case is also well-known to support a distinct conformal group. 

\m

\n{\bf Example D} The $d$-dimensional real projective space $\mathbb{RP}^k$  yet again supports an isometry group, 
\be 
Isom(\mathbb{RP}^k) = SO(k + 1)  \m , 
\ee 
but not a distinct similarity group.   
It additionally supports a projective transformation group 
\be 
Proj(\mathbb{RP}^{d - 1}) = PGL(d, \mathbb{R}) 
\ee 
as enters Image Analysis from the general and directly-realized perspective of a finitely-placed observer \cite{PE16}.  
For $\FrC^d = \mathbb{R}^d$, $\FrP^{d - 1} = \mathbb{RP}^{d - 1}$.  

\m 

\n{\bf Example E} $\mathbb{R}^d$ only supports affine and equiareal groups -- distinct from the similarity and Euclidean groups respectively -- for $d \geq 2$.

\subsection{Relational spaces: constellation spaces quotiented by automorphism groups}

\n{\bf Structure 1} One can consider a corresponding relational theory on the quotient configuration space relational space
\be 
\FrR(\FrC^d, N)  \:=  \frac{\FrQ(\FrC^d, N)}{Isom(\FrC^d)} 
                 \:=  \frac{\bigtimes_{i = 1}^N \FrC^d}{Isom(\FrC^d)}  \m . 
\ee
\n{\bf Structure 2} A given absolute space $\FrC^d$ may moreover admit generalized symmetries $Gen\mbox{-}Sym(\FrC^d)$ corresponding to geometry $\lFrg$, 
furnishing a $\lFrg$-relational theory with configuration space 
\be
\Rel(\FrC^d, N; Gen\mbox{-}Sym(\FrC^d)) \:=  \frac{\FrG(\FrC^d, N)}{Gen\mbox{-}Sym(\FrC^d)}   \m .
\ee  
More precisely, given absolute space $\FrC^d$ may admit a nontrivial (perhaps competing) bounded lattice $\lattice$ 
of geometrically meaningful automorphism groups corresponding to which members of the competing lattice of generalized Killing equations possess nontrivial solutions.  
(`Competing lattice' is meant in the sense of \cite{AMech}: of geometrically-significant subgroups of either the conformal group or of the affine group, the two being 
incompatible in Flat Geometry rather than mutually realizable therein.) 
This gives the totality of generalized relational theories on $\FrC^d$ as a package.
A relational theory is for now a triple $(\FrC^d, N; \lFrg)$ where $\lFrg$ is some geometrically meaningful automorphism group acting on $\FrC^d$.  

\m 

\n{\bf Definition 3} {\it Relational space} is the quotient space
\be  
\Rel(\FrC^d, N, Aut(\FrQ, \sigma))  \:=  \frac{\FrQ(\FrC^d, N)}{Aut(\langle\FrQ, \sigma\rangle)}       \m .  
\ee
\n{\bf Definition 4} For those $\lFrg$ that do not include a scaling transformation, 
the relational space notion specializes to the {\it shape space} notion \cite{Kendall84, Kendall, FileR, AMech, PE16} 
\be 
\FrS(\FrC^d, N; \lFrg) := \Rel(\FrC^d, N; \lFrg)   \m . 
\ee
\n{\bf Definition 5} For those $\lFrg$ that do include a scaling transformation, 
the relational space notion specializes to the {\it shape-and-scale space} notion \cite{LR95-97, FileR, AMech, ABook} 
\be 
\FrR(d, N; \lFrg) := \Rel(\FrC^d, N; \lFrg)        \m .
\ee
\n{\bf Remark 9} Relational Theory is thus a portmanteau of Shape Theory and Shape-and-Scale Theory.  
The distinction of whether or not scaling is among the automorphisms is significant in practice because many of the most-studied models are part of a 
{\it shape space and shape-and-scale-space pair}.
This corresponds to Shape Theories which remain algebraically consistent upon removal of an overall dilation generator. 
However, there are more generally plenty of instances of singletons, of which one is given below and others are listed in e.g. \cite{Project-1}.  

\m 

\n{\bf Example 0} For absolute space $\mathbb{R}^d$, quotienting out the constellation space by the similarity group $Sim(d)$ gives 
{\it Kendall's Similarity Shape Theory} \cite{Kendall84}, with corresponding shape spaces 
\be 
\FrS(d, N)  \:=   \frac{\FrQ(d, N)}{Sim(d)} 
            \es   \frac{\FrQ(d, N)}{Tr(d) \rtimes (Rot(d) \times Dil)} 
            \es   \frac{\bigtimes_{i = 1}^N \mathbb{R}^d}{\mathbb{R}^d \rtimes (SO(d) \times \mathbb{R}_+)}  
            \es   \frac{\bigtimes_{i = 1}^n \mathbb{R}^d}{SO(d) \times \mathbb{R}_+}                               \m .
\ee
\n{\bf Example 1} For absolute space $\mathbb{R}^d$, quotienting out the constellation space by the Euclidean group $Eucl(d)$ gives {\it Metric Relational Theory}.
Here,
\be 
\FrR(d, N)  \:=   \frac{\FrQ(d, N)}{Isom(\mathbb{R}^d)}
            \es   \frac{\FrQ(d, N)}{Eucl(d)} 
            \es   \frac{\FrQ(d, N)}{Tr(d) \rtimes Rot(d)} 
            \es   \frac{\bigtimes_{i = 1}^N \mathbb{R}^d}{\mathbb{R}^d \rtimes SO(d)}  
            \es   \frac{\bigtimes_{i = 1}^n \mathbb{R}^d}{SO(d)}                                                   \m .
\ee 
for $n : = N - 1$.

\m 

\n{\bf Example 2} Quotienting out by the affine group $Aff(d)$ gives {\it Affine Shape Theory}, whose shape space is  
\be 
\FrS(d, N; Aff(d))  \:=   \frac{\FrQ(d, N)}{Aff(\mathbb{R}^d)} 
                    \es   \frac{\FrQ(d, N)}{Tr(d) \rtimes GL(d, \mathbb{R})} 
                    \es   \frac{\bigtimes_{i = 1}^N \mathbb{R}^d}{\mathbb{R}^d \rtimes GL(d, \mathbb{R})}    
			        \es   \frac{\bigtimes_{i = 1}^n \mathbb{R}^d}{GL(d, \mathbb{R})}                               \m .
\ee 
\n{\bf Example 3} Quotienting out by the equi-top-form group $Equi(d)$ gives {\it Equi-top-form Scale-and-Shape Theory}, whose scale-and-shape space is  
\be 
\FrR(d, N; Equi(d))  \:=   \frac{\FrQ(d, N)}{Equi(\mathbb{R}^d)} 
                     \es   \frac{\FrQ(d, N)}{Tr(d) \rtimes SL(d, \mathbb{R})} 
                     \es   \frac{\bigtimes_{i = 1}^N \mathbb{R}^d}{\mathbb{R}^d \rtimes SL(d, \mathbb{R})}    
			         \es   \frac{\bigtimes_{i = 1}^n \mathbb{R}^d}{SL(d, \mathbb{R})}                              \m .
\ee 
\n{\bf Example 4} Quotienting out by the conformal group $Conf(d)$ gives {\it Conformal Shape Theory}, whose shape space is  
\be 
\FrC(d, N)  \:=   \frac{\FrQ(d, N)}{Conf(\mathbb{R}^d)}  
            \es   \frac{\bigtimes_{i = 1}^N \mathbb{R}^d}{SO(d, 1)}          \m .
\ee 
\n{\bf Example 5} With the circle $\mathbb{S}^1$ in the role of absolute space, 
quotienting out the constellation space by the corresponding isometry group gives {\it Metric Shape-and-Scale Theory}, with scale-and-shape space  
\be 
\FrR(\mathbb{S}^1, N)  \:=   \frac{\FrQ(\mathbb{S}^1, N)}{Isom(\mathbb{S}^1)}
                       \es   \frac{\bigtimes_{i = 1}^N \mathbb{S}^1}{\mathbb{S}^1}  
                       \es   \bigtimes_{i = 1}^n \mathbb{S}^1
                       \es   \mathbb{T}^n	                         				                               \m .
\ee 
\n{\bf Example 6} For tori $\mathbb{T}^d$ in the role of absolute space, quotienting out the constellation spaces by the corresponding isometry groups 
\be 
Isom(\mathbb{T}^d)  \es  \coprod_{a = 1}^d U(1)                                                                    
                    \es  \coprod_{a = 1}^d \mathbb{S}^1 
					\es  \mathbb{T}^d 
\ee
give {\it Metric Shape-and-Scale Theory}, with scale-and-shape space  
\be 
\FrR(\mathbb{T}^d, N)  \:=   \frac{\FrQ(\mathbb{T}^d, N)}{Isom(\mathbb{T}^d)}
                       \es   \frac{\bigtimes_{i = 1}^{N \, d} \mathbb{S}^1}{\bigtimes_{i = 1}^{d} \mathbb{S}^1}  
                       \es   \bigtimes_{i = 1}^{n \, d} \mathbb{S}^1
                       \es   \mathbb{T}^{n \, d}                    	                         				  \m .
\label{Cancel}
\ee 
\n{\bf Example 7} For spheres $\mathbb{S}^d$ in the role of absolute space, quotienting out the constellation spaces by the corresponding isometry groups 
\be 
Isom(\mathbb{S}^d) = SO(d + 1)                                                                                   \m . 
\ee
give {\it Metric Shape-and-Scale Theory}, with scale-and-shape space  
\be 
\FrR(\FrS^d, N)  :=  \frac{\FrQ(\mathbb{S}^d, N)}{Isom(\mathbb{S}^d)}
                 :=  \frac{\bigtimes_{i = 1}^N \mathbb{S}^d}{SO(d + 1)}                                          \m .
\ee
So e.g.\ the $d = 2$ case's $SO(3)$ admits a $SO(2) = U(1)$ subgroup for the merely axisymmetrically-identified configurations.

\m 

\n{\bf Example 8} For $\FrC^d = \mathbb{RP}^d$, $Isom(\mathbb{RP}^d) = SO(d)$ as well.
The main example of interest is however the projective case, for which the automorphisms are
\be 
Proj(\mathbb{RP}^{d - 1}) = PGL(d, \mathbb{R})                                                                        \m .
\ee 
The relational space is then \cite{MP05}
\be 
\FrS(\mathbb{RP}^{d - 1}, N; PGL(d, \mathbb{R}))  \es  \frac{\bigtimes_{I = 1}^N \mathbb{RP}^d}{PGL(d, \mathbb{R})}   \m .
\ee

\subsection{Simple explicit manifold topology and geometry examples}\label{2.last}

\n{\bf Example 1} For Similarity Shape Theory in 1-$d$, straighforwardly 
\be 
\FrS(N, 1) = \mathbb{S}^{n - 1}  \m \mbox{(spheres)} \m 
\ee 
both topologically and metrically.  
This moreover generalizes to 
\be 
\FrP(N, d) = \mathbb{S}^{n \, d - 1}
\ee 
-- Kendall's preshape sphere -- being the result of quotienting out translations and dilations from constellation space; thus Kendall's similarity shape space itself is 
\be 
\FrS(N, d) \es \frac{\mathbb{S}^{n \, d - 1}}{SO(d)}  \m . 
\ee 
This moreover does not in general simplify. 

\m 

\n{\bf Example 2} For Similarity Shape Theory in 2-$d$, however, \cite{Smale70}
\be 
\FrS(N, 2) \es \frac{\mathbb{S}^{n \, d - 1}}{SO(2)} \es  \frac{\mathbb{S}^{n \, d - 1}}{\mathbb{S}^1} \es \mathbb{CP}^{n - 1} \m \mbox{(complex-projective spaces)} \m 
\ee
by the generalized Hopf map.
In its realization as a shape space, this is moreover equipped with the standard Fubini--Study metric \cite{Kendall84}. 
For $N = 3$, this furthermore reduces to the {\it shape sphere} of triangles, by the topological and geometrical coincidence 
\be
\mathbb{CP}^1 = \mathbb{S}^2 \m .  
\ee 
For $N = 3$ in 3-$d$, the working shares many similarities but the outcome is a shape hemisphere with edge \cite{A-Monopoles}. 

\m

\n{\bf Example 3} For Euclidean Shape-and-Scale Theory in 1-$d$,  
\be
\FrR(N, 1) = \mathbb{R}^{n} \m .
\ee 
These are the topological and metric cones over the corresponding shape spaces, a result which holds in general for Euclidean shape-and-scale spaces \cite{FileR}.

\subsection{Subsequent Relational Mechanics}\label{2.RPM}

\n{\bf Remark 10} The premise here is to view dynamics as a geodesic on the corresponding configuration space geometry. 

\m

\n{\bf Remark 11} See \cite{BB82, FileR} for Example 0)'s, and  
                     \cite{B03, FileR} for Example 1)'s, 
	                 \cite{AMech} for Example 2), 3) and 4)'s, 
	                 \cite{ASphe} for Example 5) and 7)'s and \cite{Forth} for Example 6)'s corresponding whole-universe theories of Relational Mechanics.   
\cite{LR95-97} and \cite{M02-M05} furthermore consider the role of the shape (hemi)sphere of triangles in the context of subsystem models.  

\m 

\n In Example 0)'s more habitually considered flat-space Euclidean case of Mechanics, a further substanital issue is whether to excise the maximal collision in $d \geq 2$. 
In $d = 3$, some works furthermore excise the collinear configurations, which, for $N = 3$, amount to the planar edge of the shape hemisphere.
We will comment further on these configurations and practices below, noting that they are moreover prototypical for further significant such in the wider range of 
Mechanics Theories allused to above.

\section{Strata}\label{Strata}

\n For further cases than Sec \ref{2.last}, we need more mathematics, as follows.

\m 

\n{\bf Global Problem 1} Quotienting, as enters relational approaches through how these handle Configurational Relationalism.  
{\sl Reductions} 
\be 
\FrQ \m \longrightarrow \m \widetilde{\FrQ}   \es  \frac{\FrQ}{\lFrg}
\ee 
{\sl usually kick one out of the class of manifolds into the class of stratified manifolds}.  

\m 

\n{\bf Remark 1} Stratified manifolds are somewhat unfamiliar to much of the Theoretical Physics community and so require some explaining. 
A first point of order is that, while these are not topologcial {\sl manifolds}, they are still topological {\sl spaces}, 
so we start by recollecting this standard albeit rather less structured notion.

\subsection{Topological spaces} 

\n{\bf Definition 1} A {\it topological space} \cite{Sutherland} is a collection $\tau$ of open subsets $\FrU_O$ of a given set $\FrX$ with the following properties. 

\m 

\n{\bf Topological Space 1)} $\FrX, \emptyset \in \tau$.

\m 

\n{\bf Topological Space 2)} The union of any collection of the $\FrU_O$ is also in $\tau$.

\m 

\n{\bf Topological Space 3)} The intersection of any finite number of the $\FrU_O$ is also in $\tau$.  

\m 

\n{\bf Remark 2} Below we build up an account of some topological properties which enter our discussion \cite{Wald}.   

\m 

\n{\bf Definition 2} A collection of open sets $\{\FrU_{C}\}$ is an {\it open cover} for $\FrX$ if $\FrX = \bigcup_{C} \FrU_{C}$. 

\m 

\n On the one hand, a subcollection of an open cover that is still an open cover is termed a {\it subcover}, $\{\FrV_{D}\}$ for $\fD$ a subset of the indexing set $\fC$. 

\m 

\n On the other hand, an open cover $\{\FrV_{D}\}$ is a {\it refinement} of $\{\FrU_{C}\}$ 
if to each $\FrV_{D}$ there corresponds a $\FrU_{C}$ such that $\FrV_{D} \subset \FrU_{C}$.  
$\{\FrV_{D}\}$ is furthermore {\it locally finite} if each $x \in \FrX$ has an open neighbourhood $\FrN_x$ such that only finitely many 
$\FrV_{D}$ obey $\FrN_x \bigcup \FrV_{D} \neq \emptyset$.  

\m

\n{\bf Definition 3} A topological space $\tau(\FrX)$ is {\it compact} \cite{Armstrong, Sutherland, Lee1} if every open cover of $\FrX$ has a finite subcover. 

\m 

\n{\bf Remark 3} Compactness is useful e.g.\ through its generalizing continuous functions being bounded on a closed interval of $\mathbb{R}$.

\m

\n{\bf Definition 4} A topological space $\tau(\FrX)$ is {\it paracompact} \cite{Lee1} if every open cover of $\FrX$ has a locally finite refinement. 

\m 

\n{\bf Remark 4} Paracompactness is useful e.g.\ through its permitting use of partitions of unity \cite{Munkres, Lee2}. 

\m 

\n{\bf Remark 5} Notions of {\it separation} are topological properties which indeed involve separating two objects 
(points, certain kinds of subsets) by encasing each in a disjoint subset.  

\m 

\n{\bf Definition 5} {\it Hausdorffness} \cite{Sutherland, Lee1} is a particular such, for which  
$$
\mbox{for } \m \mbox{$x \mma y \in \FrX, x \neq y \mma  \exists$ \m open sets \m ${\FrU}_x \mma {\FrU}_y \in \tau$}
$$
\beq
\mbox{such that \m $x \in {\FrU}_x \mma y \in {\FrU}_y$ and ${\FrU}_x \bigcap {\FrU}_y = \emptyset$} \m . 
\label{Hausdorffness}
\eeq
I.e.\ separating points by open sets.  

\m 

\n{\bf Remark 6} Hausdorffness allows for each point to have a neighbourhood without stopping any other point from having one.
This is a property of the real numbers, and is additionally permissive of much Analysis.
In particular, Hausdorffness secures uniqueness for limits of sequences. 
%

\m 

\n{\bf Remark 7} Some notions of countability are concurrently topological properties, due to involving counting of topologically defined entities \cite{Lee1}.

\m

\n{\bf Definition 6} {\it First countability} holds if for each $x \in \FrX$, 
there is a countable collection of open sets such that every open neighbourhood $\FrN_x$ of $x$ contains at least one member of this collection. 

\m

\n{\bf Definition 7} {\it Second countability} is the stronger condition that 
there is a countable collection of open sets such that every open set can be expressed as union of sets in this collection.   
Second-countability is also useful via being a property standardly attributed to manifolds.  

\m 

\n{\bf Definition 8} A topological space $\langle \FrX, \tau \rangle$ is {\it locally Euclidean} if every point $x \in \FrX$ has a neighbourhood $\FrN_x$ 
that is homeomorphic to $\mathbb{R}^p$: Euclidean space.

\subsection{Topological manifolds} 

\n{\bf Remark 8} Topological manifolds are topological spaces blessed with the combination of being Hausdorff, second-countable and locally Euclidean.  
This combination happens to imply paracompactness as well \cite{Munkres, Lee2}.  
Local Euclideanness moreover underlies the use of charts in the study of differentiable manifolds.  

\m  

\n{\bf Remark 9} Hausdorff and second-countability is good for analytical tractability, in the manner of a balance point. 
In some sense, topological spaces which re not Hausdorff have too few open sets for much of Analysis, 
whereas second-countable ones have too many, yet ones which have both are `just right'. 
So, much as Goldilocks would discard Daddy Bear's porridge for not being sweet enough and Mummy Bear's porridge for being too sweet  
-- in favour of Baby Bear's porridge which is `just right' -- Analysts would much prefer to work with topological spaces which are both Hausdorff and second-countable.  

\m 

\n{\bf Remark 10} Our first issue however corresponds to the loss of the locally-Euclidean pillar of manifoldness in the case of stratified manifolds. 
On the one hand, this loss corresponds to a stratified manifold in general having {\sl multiple} dimensions, in a piecewise manner.
This is moreover only a {\sl local} loss of local Euclideanness, and there are fairly benevolent rules for `patching together' of the pieces
as put forward by Whitney and Thom \cite{W-T, Whitney65, Thom69}.

\subsection{Stratified manifolds} 

\n{\bf Structure 1} Let $\FrX$ be a topological space that can be split according to 
\be
\FrX = \FrX_{\sp} \cup \FrX_{\sq}                                                                                                          \m .
\ee 
for which 
\be 
\mbox{dim}_{\sp}(\FrX) = \mbox{dim}(\FrX)
\ee 
and 
\be 
\FrX_{\sq} := \FrX - \FrX_{\sp}                                                                                                            \m .
\ee  
Next consider recursive such splittings, so e.g.\ $\FrX_{\sq}$ further splits into $\{\FrX_{\sq}\}_{\sp}$ and $\{\FrX_{\sq}\}_{\sq}$.
Then set
\be 
\FrM_1 = \FrX_{\sp} \mma \FrM_2 = \{\FrX_{\sq}\}_{\sp} \mma \FrM_3 = \{\{\FrX_{\sq}\}_{\sq}\}_{\sp}  \m ...
\ee 
to obtain  
\be 
\FrX = \FrM_1 \cup \FrM_2 \cup  \, ... \,  , \mbox{dim}(\FrX) = \mbox{dim}(\FrM_1) \, > \, \mbox{dim}(\FrM_2) \, > \m ...                  \m ,
\ee
where each $\FrM_{I}$, $I = 1, 2, \, ...$ is itself a manifold.
This provides a partition of $\FrX$ by dimension. 
$\FrX$ is moreover only a topological manifold in the case of a trivial (i.e.\ single-piece) partition.

\m 

\n{\bf Definition 9} A {\it strict} partition of a topological space is a (locally finite) partition into strict manifolds. 
A manifold $\FrM$ within a $m$-dimensional open set $\FrW$ is $\FrW${\it -strict} if its $\FrW$-{\it closure} 
\be
\overline{\FrM}  :=  \FrW - \mbox{Clos}\,\FrM
\ee 
and the $\FrW$-{\it frontier} 
\be 
\FrM^{\sF}       :=  \overline{\FrM} - \FrM
\ee 
are topological spaces in $\FrW$. 

\m 

\n{\bf Definition 10} A set of manifolds in $\FrW$ has the {\it frontier property} if, for any two distinct such, say $\FrM$ and $\FrM^{\prime}$,  
\beq
\mbox{ if }   \m  \FrM^{\prime} \cap \overline{\FrM} \m \neq \m  \emptyset \mma 
\mbox{ then } \m  \FrM^{\prime} \subset \overline{\FrM}              \m \mbox{ and } \m \m 
                  \mbox{dim}(\FrM^{\prime}) \, <  \, \mbox{dim}({\FrM}) \mbox{ } .
\label{Frontier}
\eeq
A partition into manifolds itself has the frontier property if the corresponding set of manifolds does.  

\m 

\n{\bf Definition 11} A {\it stratification} of $\FrX$ \cite{Whitney65} is a strict partition of $\FrX$ that possesses the frontier property. 
The corresponding set of manifolds are known as the {\it strata} of the partition.

\m

\n{\bf Definition 12} The {\it principal stratum} is the one whose corresponding {\sl orbit} is of minimal dimension.

\m 

\n{\bf Remark 11} Stratified manifolds have additionally been equipped with differentiable structure \cite{Thom69} (see e.g.\ \cite{Pflaum} for a modern account).

\subsection{Conceptual classification of stratified manifolds} 

\n{\bf Remark 12} The following conceptual classification of stratified manifolds is useful in subsequent sections.  

\m

\n{\bf Type i) Trivially-contiguous}. 

\m 

\n These stratified spaces can be qualified as manifolds with boundaries, corners \cite{Lee1, Lee2} (etc.\ in higher dimensions), 
in which some of the boundaries, corners etc are geometrically distinct, but are still contiguous to the top stratum in the manner of manifold theory.

\m 

\n{\bf Example 1)} Flat-space Euclidean Shape-and-Scale Theory in $d \geq 2$ has a distinct-stratum maximal collision O positioned as a cone apex. 
For $d \geq 3$ non-collision collinearities C form a separate stratum, e.g.\ the bounding plane of the $N = 3$ shape hemisphere, minus one puncture point: O itself.  
This rests on the 3-$d$ rotation group $SO(3)$ acting fully on the generic configurations G, but $SO(2)$ being inactive on collinear configurations C, 
and the entirety of $SO(3)$ being inactive on O. 
In contrast, for $d = 2$, the 2-$d$ rotation group $SO(2)$ acts no differently -- fully -- on G and C, while the entirety of $SO(2)$ is inactive acting on O.  
This underlies 3 particles in 2-$d$ being a {\sl mathematically distinct} model from 3 particles in 3-$d$.  

\m

\n{\bf Type ii) Topologically-nice nontrivially-contiguous}, this contiguity lying within the remit of stratified manifolds' axioms.    
By `topological niceness' we mean one of the following holds. 

\m 

\n a) Either that $\FrX$ be either LCHS (locally compact Hausdorff second-countable) \cite{Lee1}. 

\m 

\n b) Or that $\FrX$ be LCHP (locally compact Hausdorff paracompact) \cite{Munkres}; 
there is moreover considerable degeneracy between paracompactness and second-countability in the current context.

\m

\n I.e.\ 2 of the 3 pillars of manifoldness are kept and moreover supplemented with the following further analytic niceness condition. 

\m 

\n{\bf Definition 13} A topological space $\FrX$ is {\it locally compact} \cite{Lee1} if each point $\mp \in \FrX$ is contained in a compact neighbourhood.    

\m

\n{\bf Remark 13} LCHS and LCHP are moreover standard and well studied-packages with applications in many further areas of Mathematics. 

\m 
 
\n{\bf Remark 14} Conceptual and computational schemes have furthermore been provided in the specific case of stratified manifolds which are LCHS by Kreck \cite{Kreck} 
                                                                                                                                         and LCHP by Pflaum \cite{Pflaum}.

\m 

\n{\bf Type iii) Topologically-complex nontrivially-contiguous}. 

\m 

\n{\bf Remark 15} Quotienting a topological space by an equivalence relation, $\langle \FrX, \tau\rangle/\,\widetilde{\mbox{ }}$, 
produces the corresponding {\it quotient topology} \cite{Lee1, Munkres}.

\m 

\n{\bf Remark 16} This moreover does not preserve some topological properties; this applies in particular to all three manifoldness properties. 
We focus particularly on failure of quotients to be Hausdorff-separable; this requires first contemplating what form weaker notions of separation take.  

\m 

\n{\bf Definition 14} A topological space $\FrX$ is {\it Fr\'{e}chet} \cite{Willard} alias accessible if whichever 2 distinct points $x \neq y$ in $\FrX$ are separated 
by there being 
\be 
\mbox{at least one open set $\FrU$ such that $x \in \FrU$ but $y \not{\hspace{-0.05in}\in} \m \FrU$} \m . 
\ee 
\n{\bf Remark 17} Contrast with Hausdorffness, for which {\sl two distinct} open sets are involved in a symmetric manner. 

\m

\n{\bf Definition 15} A topological space $\FrX$ is {\it Kolmogorov} \cite{Willard} if whichever 2 distinct points $x \neq y$ in $\FrX$ are topologically distinguishable, 
meaning that there is  
\be 
\mbox{at least one open set \m $\FrU$ \m  such that \m $x \in \FrU$ \m but \m $y \not{\hspace{-0.05in}\in} \m \FrU$}   \mma  
                                  \mbox{or } \m  \mbox{$x \in \FrU$ \m  but \m $y \not{\hspace{-0.05in}\in} \m \FrU$}  \m .
\ee 
\n{\bf Remark 18} There is moreover a construct by which Kolmogorovness is guaranteed as a minimum standard of separability. 

\m

\n{\bf Example 2} In the Affine Shape Theory of quadrilaterals in the plane, both the collinear and generic shapes form their own real projective space $\mathbb{RP}^2$ stratum, 
\be 
\mathbb{RP}^2 \m \disjoint \m \mathbb{RP}^2           \m , 
\ee 
with every collinear configuration C lying arbirtarily close to every generic configuration G.
This effect rests on $GL(d, \mathbb{R})$ possessing a $SO(d)$ subgroup, with $(d, N) = (2, 4)$ being minimal to manifest this effect.  
This impossibilitates Hausdorff or even Fr\'{e}chet separability as Groisser and Tagare have shown \cite{GT09}; 
all that one is left with is the much weaker and {\sl qualitatively distinct} Kolmogorov separability. 

\m

\n{\bf Remark 19} Theorems of Analysis are more sparsely available here, and computational schemes for stratified manifolds of this more general nature remain to be developed.  
Thus substantial technical problems are posed by this affine example.
Note moreover that Kendall-Type Relational Theories have not realized Type ii), by which the divide into Types i) and iii) is marked; 
we will have more to say about this in the Conclusion.   

\m
 
\n{\bf Example 3} While Projective Geometry is well-known to cure Affine Geometry of various imperfections, 
passing to Projective Shape Theory {\sl does not} remove the corresponding Shape Theory's shape space's mere Kolmogorov-separatedness.
This was recently demonstrated by Kelma, Kent and Hotz \cite{KKH16}.  

\m 

\n{\bf Global Problem 2} Is that merely-Kolmogorov separated stratified manifolds can arise from configuration space reductions.

\subsection{Strategies for handling stratification} 

\n{\bf Strategy A)} {\it Excise Strata}. This consists of discarding all bar the principal stratum.

\m 

\n{\bf Strategy B)} {\it Unfold Strata}. Here non-principal strata are unfolded so as to end up possessing the same dimension as the principal stratum.   

\m

\n{\bf Strategy C)} {\it Accept All Strata}. 

\m 

\n{\bf Remark 20} While Excise Strata simplifies the remaining mathematics to handle, it can be a crude approximation and an unphysical manoeuvre.
Excising strata is e.g.\ often used in the context of removing collinearities from the 3-$d$ $N$-body problem, or of maximal collisions in 2- or 3-$d$.  

\m 

\n{\bf Remark 21} Incidentally, Example 2 above illustrates a further problem with excising non-principal strata: here the nonprincipal stratum is of {\sl equal dimension} 
with the principal stratum.  

\m 

\n{\bf Remark 22} Unfold Strata was considered e.g.\ by Fischer \cite{Fischer86}.       
One may however then question whether such an unfolding is physically meaningful and mathematically unique.    

\m 

\n{\bf Remark 23} Prima facie, Accept All Strata is the strategy which is accord with Leibniz's Identity of Indiscernibles \cite{L}.  

\m 

\n{\bf Remark 24} On the one hand, Excise Strata and Unfold Strata remain within the familiar and mathematically tractable remit of Manifold Geometry and Fibre Bundles thereover.
On the other hand, Accept all Strata harder mathematics being required: Fibre Bundles do not suffice due to heterogeneity amongst what might have elsewise been fibres.  
To handle this, one needs at least general bundles \cite{IshamBook2, Husemoller}, and, for a wider range of applications, sheaves. 
\cite{ABook}'s relational program favours Accept All Strata, arguing against excision of collinearities and maximal collisions, Fischer's unfolding construct \cite{Fischer86}, 
and the excision aspect of the remit of Bartnik and Fodor's Thin Sandwich Theorem \cite{BF}. 

\m 

\n{\bf Remark 25} The idea of geometry as a geodesic on configuration space becomes implemented by a geodesic on a stratified manifold. 
While geodesics can be defined locally tratum-by-stratum therein, 
one would also wish to know what happens whenever such geodesics strike boundaries between strata, 
and this remains unclear; let us term this {\bf Global Problem 3}.  
In Type iii) strata, moreover, there is the problem that all points in some of the most significant strata are arbitrarily close to all points in other strata,
so the significance of strata by strata geometrical structure, including of geodesics, becomes obscured.  

\m  

\n{\bf Remark 26} We will moreover be considering a fourth strategy once we have discussed the GR counterpart.

\subsection{Sheaf Methods}\label{Cl-Glob-Sheaves}

\n{\bf Structure 2} Fibre bundles extend study of a given manifold by attaching identical fibres to each of its points, forming a total space;
points on each fibre moreover project down onto each of the original (`base') manifolds' points.
Some global properties of manifolds follow from considerations of sections: cuts through the total space to which each fibre contributes just one point.  

\m 

\n{\bf Remark 27} Whereas the fibres attached to each base space point within a given fibre bundle are all the same, sheaves allow for heterogeneous attached objects.   
Indeed, imagine a variety of shapes and sizes of `grain' attached to a `stem', in analogy with a `sheaf of wheat'. 
This situation is often encountered in the study of stratified manifolds, by which fibre bundle theory has ceased to suffice ({\bf Global Problem 4}).  

\m 

\n{\bf Structure 3} While {\it general bundles} \cite{IshamBook2, Husemoller} already admit this required feature, 
working in terms of sheaves offers further advantages in global methodology. 
First of all, sheaves are based on a mathematical reconceptualization in which restriction maps play a central role.
Sheaves moreover posses an analogue of fibre bundles' key global notion of section; 
a sheaf-level analogue of the Gribov effect \cite{HTBook} consequently appears, and sheaf-level analogues of topological defect theory is expected as well ({\bf Global Problems 5 and 6}).
The advantages sheaves furthermore offer derive from their possessing two additional notions: a `local to global' gluing and a `global to local' condition.  
{\sl Sheaves are tools for tracking locally defined entities by attaching them to open sets within a topological space.}
We need not further detail sheaves for the purposes of the current paper; see \cite{ABook} for an outline and \cite{Sheaves} for detailed texts.

\m 

\n{\bf Remark 28} The above notions can indeed be applied to stratified configuration spaces $\FrQ$ 
(and to the corresponding stratified phase spaces as well \cite{2000-Proceedings}).

\m 
 
\n{\bf Application 1} Sheaves can be used in meshing together the heterogeneous types of charts possessed by a stratified manifold.

\m 
 
\n{\bf Application 2} Sheaves can be used to define metric-level geodesics within stratified manifolds, 
i.e.\ the situation arising from viewing dynamics as geodesics on configuration spaces, once reduced such become in general stratified...
This sheaf approach points toward handling paths that move into boundaries between strata, and thus e.g.\ to geodesic principles upon stratified manifolds. 

\m 
 
\n{\bf Application 3} Thirdly, Kreck's {\it stratifold} \cite{Kreck} consists of (stratified manifold, sheaf) pairs in the case of LCHS stratified manifolds.   

\m 

\n{\bf Application 4} Tangent, cotangent, symplectic and Poisson spaces, 
in each case corresponding to stratified configuration spaces can be studied using Sheaf Methods \cite{2000-Proceedings}.
Sheaf Methods have also started to be applied to gauge orbit spaces has also begun \cite{2000-Proceedings}.

\m 

\n We finally give a partial answer to one of the 120 foundational questions posed in \cite{ABook}.  

\m 

\n{\bf Research Project 41)} ``{\sl To what extent can Sheaf Methods advance our understanding of $N$-body Problem configuration and phase spaces?}"  

\m 

\n{\bf Answer} In Sheaf Methods' current state of development, the answer is yes for collinearities, no for maximal collisions and no for Affine or Projective Shape Theory examples.

\section{General Relativity configuration spaces}

\subsection{Incipient configuration space $\Riem(\bupSigma)$} 

\n We next outline the GR counterpart of these relational considerations, working in dynamical formulation with `traditional' geometrodynamical variables \cite{ADM, DeWitt67}.    

\m 
																																	 
\n{\bf Remark 1} GR's incipient configurations are Riemannian 3-metrics $\bh$ with components $\mh_{ab}(x^c)$ (for $x^c$ coordinates in space) 
                 on a fixed 3-topology $\bupSigma$ interpreted as a spatial slice of spacetime 
				 (itself a semi-Riemannian 4-metric $\bg$ with components $\mg_{\mu\nu}$ on a 4-topology $\FrM$). 

\m 

\n{\bf Remark 2} We consider compact without boundary models for space $\bupSigma$, with $\mathbb{S}^3$ and $\mathbb{T}^3$ the most commonly considered specific spatial topologies.
				   
\m

\n{\bf Remark 3} As the 3-metric $\bh$ is a symmetric $3 \times 3$ matrix, it has 6 degrees of freedom per space point. 

\m  

\n{\bf Definition 1} The totality of GR's $\bh$ on a fixed $\bupSigma$ constitute GR's incipient configuration space $\Riem(\bupSigma)$.  

\m 

\n{\bf Remark 4} $\Riem(\bupSigma)$ is supplied with its own metric by GR's action in split space-time form \cite{ADM}: the inverse DeWitt metric \cite{DeWitt67}.

\subsection{Diffeomorphims $Diff(\bupSigma)$ in role of automorphisms} 

\n{\bf Structure 1} The spatial diffeomorphisms $Diff(\bupSigma)$ are GR's automorphism group acting on the incipient configurations regarded as physically meaningless.
This is a differential-geometric level endeavour.

\m 

\n{\bf Remark 5} This has infinitesimal implementation by Lie derivative, totalling 3 redundant degrees of freedom per space point.

\subsection{$\lSuperspace(\bupSigma)$ as a further relational space} 

\n{\bf Definition 2} 3-Metric $\bf$ minus 3-diffeomorphism information leaves 3-geometric information $\bG^{(3)}$: 
a notion of scaled shape, analogous to Euclidean Relationalism's rotationally-invariant inner products. 

\m 

\n{\bf Remark 6} Counting out, this has $6 - 3 = 3$ degrees of freedom. 

\m 

\n{\bf Structure 2} Wheeler's \cite{Battelle, DeWitt67, Giu09} 
\be
\Superspace(\bupSigma) \es  \frac{\Riem(\bupSigma)}{Diff(\bupSigma)} \m  .  
\label{Intro-Superspace}
\ee

\subsection{Conformal transformations as further automorphisms} 

\n{\bf Structure 3} Local scaling $Conf(\bupSigma)$ of spatial conformal transformations.  
This has 1 degree of freedom per space point.

\subsection{$\CRiem(\bupSigma)$ and $\CS(\bupSigma)$ as further relational spaces} 

\n{\bf Remark 7} The corresponding invariants are unit-determinant metrics 
\be
\buu  \es  \frac{\bh}{^3\sqrt{\mh}}
\ee
for $\mh := \mbox{det} \, \bh$.  

\m 

\n{\bf Structure 4} The space of these is {\it conformal Riem} 
\be 
\CRiem(\bupSigma)  \es  \frac{\Riem(\bupSigma)}{Conf(\bupSigma)}    \m . 
\ee 
This has $6 - 1 = 5$ degrees of freedom per space point.
See Chapter 21 of \cite{ABook} for a conceptual discussion of this space and \cite{DeWitt67, DeWitt70, FM96} for more technical discussion.
This is GR's counterpart of Kendall's preshape sphere, 
including as regards being relatively simpler topologically and geometrically than the corresponding theory's other reduced configuration spaces.  

\m  

\n{\bf Remark 8} We next consider 
\be 
Conf(\bupSigma) \rtimes Diff(\bupSigma)
\label{CD}
\ee 
in the role of automorphism group; c.f.\ $Eucl(d) = Tr(d) \rtimes Rot(d)$ or $Dil \times Rot(d)$ in Mechanics models. 
This has $6 - 3 - 1 = 2$ degrees of freedom per space point.
%
%
\n The corresponding invariants are now conformal 3-geometries $\bC^{(3)}$ \cite{York74}.  

\m 

\n{\bf Structure 5} The space of these is York's \cite{York74} {\it conformal superspace} 
\be
\CS(\bupSigma) = \frac{\Riem(\bupSigma)}{Conf(\bupSigma) \rtimes Diff(\bupSigma)} \m ,  
\ee 
further studied in \cite{FM96, ABFKO, BO10}.

\subsection{$\lSuperspace(\bupSigma)$ and $\CS(\bupSigma)$ stratification} 

\n For GR, it is $\bh$ with nonzero Killing vectors -- spatial metrics with symmetries -- which form the nontrivial stratification.  
DeWitt gave a conceptual account of this in \cite{DeWitt70}, with Fischer concurrently providing a superbly detailed technical account \cite{Fischer70}.  

\m 

\n Strata occur here, corresponding to metrics possessing Killing vectors in $\Superspace(\bupSigma)$ or conformal Killing vectors in $\CS(\bupSigma)$.  

\m 

\n One distinction with the Relational Mechanics case is that these GR configuration spaces realize \cite{Fischer70} an `inverse frontier condition' 
rather than the above-stated frontier condition.  

\m 

\n{\bf Partial answer} (to Research Project 41) While some stratifolds based on infinite-dimensional function spaces have already been contemplated \cite{Ewald-Tene-KT15}, 
the current state of development of these is not yet ready to tackle GR $\Superspace(\bupSigma)$ or $\CS(\bupSigma)$.

\subsection{The generic-$\bupSigma$ case} 

\n Fischer and Moncrief \cite{FM96} pointed out that the generic 3-manifold $\bupSigma$ having no Killing vectors means that its isometry group is trivial. 
Consequently the isotropy group is trivial, so no stratification arises from quotienting out the 3-diffeomorphisms, so 
\be
\Superspace(\bupSigma) \m \mbox{ remains a manifold}   \m .  
\ee
A similar argument applies moreover to the conformal case \cite{FM96}, evoking conformal genericity, Killing vectors, isometries, and quotienting out (\ref{CD}), so 
\be
\CS(\bupSigma) \m \mbox{ remains a manifold}           \m  
\ee
as well.  

\m 

\n So one approach to Relationalism in GR is to allow for `a metrically small and topologically nontrivial twist' in one's spatial model. 

\m 

\n Strategy D) {\it Unfold Strata Purely by Enhanced Physical Modelling}.  

\m
 
\n{\bf Further answer} (to Research Project 41) In the GR counterpart, if one works with generic $\bupSigma$, one encounters no isometries, 
thus no isotropy groups, thus no strata, and thus no need for Sheaf Methods to replace fibre bundle considerations.  
One is however unable to proceed with almost any standard GR calculations, out of almost all of these being rooted \cite{MacCallum} in symmetric rather than generic $\bupSigma$ 
(or expansions thereabout).

\section{Relational Theory (and so Shape Theory) are trivial for generic absolute space} 

\subsection{Developing a suitable notion of genericity for Kendall-Type Relational Theories}

\n{\bf Definition 1} A common sense of {\it manifold genericity} is 
\be
Isom(\FrM^d) = id                                                            \m .
\ee 
\n We now interpret this moreover more specifically as {\it metric genericity}.

\m

\n{\bf Remark 1} We additionally rephrase it as 
\be 
\mbox{Ker}{\cK} = \mbox{Ker}\cK(\FrC^d, \bh) = 0                            \m .  
\ee 
\n{\bf Remark 2} Metric genericity however does not suffice for our purpose of characterizing manifolds supporting no nontrivial relational theories. 

\m 

\n{\bf Definition 2} Our first generalization is to {\it geometric genericity}, 
\be 
\mbox{all } \m Gen\mbox{-}Isom(\FrC^d) = id                             \m . 
\ee
\n{\bf Remark 3} This is determined by  
\be
\mbox{Ker}(\cK(\FrC^d, \sigma)) = 0 \m \m \forall \m \sigma \, \in \, \lattice
\ee  
i.e.\ that the entire competing lattice of generalized Killing equations that $\FrC^d$ supports have but trivial kernels.

\m 

\n{\bf Remark 4} This generalization does not moreover cover how Projective Shape Theory is assigned. 
Including this case requires that 
\be 
Proj(\FrP^{d - 1}) = id \m  
\ee 
as well, where $\FrP^{d - 1}$ is the {\it projectivized version} of the incipient carrier space. 

\m 

\n{\bf Remark 5} This extra condition can be rephrased as 
\be 
\mbox{Ker}\cK(\FrP^d, \bP) = 0                            \m , 
\ee 
for $\bP$ the projective level of structure on $\FrP^{d - 1}$.  

\m 

\n{\bf Remark 6} For good measure, symmetry, and anticipation that partial realizations of Relationalism on the projectivized version of carrier space will attain significance, 
we demand moreover that 
\be
\mbox{Ker}(\cK(\FrP^d, \sigma)) = 0 \m \m \forall \m \sigma \, \in \, \Pi
\ee 
for $\Pi$ the lattice of geometrically-significnat subgroups of $Proj(\FrP^{d - 1})$. 

\m 

\n{\bf Definition 3} This determines the condition of {\it geometric genericity of carrier space} $\FrC^d$ {\sl and its projectivization} $\FrP^{d - 1}$.  

\m 

\n{\bf Remark 7} In the context of finite-particle Relational Theory, it is this which we subsequently refer to as genericity.

\subsection{Triviality of Kendall-Type Relational Theories on generic carrier space}

\n{\bf Proposition 1} For $\FrC^d$ generic,
\be 
\Rel(\FrC^d, N; Gen\mbox{-}Isom(\FrC^d))  \es  \frac{\FrQ(\FrC^d, N)}{Isom(\FrC^d)} 
                                          \es  \frac{\bigtimes_{I = 1}^{N}  \FrC^d}{id}
							              \es  \bigtimes_{I = 1}^{N}        \FrC^d                                   \m , 
\ee 
and 
\be 
\Rel(\FrP^{d - 1}, N; Gen\mbox{-}Isom(\FrP^{d - 1}))  \es  \frac{\FrQ(\FrP^{d - 1}, N)}{Isom(\FrP^{d - 1})} 
                               \es  \frac{\bigtimes_{I = 1}^{N}  \FrP^{d - 1}}{id}
							   \es  \bigtimes_{I = 1}^{N}        \FrP^{d - 1}                                          
\ee 
are the only options available for relational spaces, and corresponding relational theories.
I.e.\ just the constellation space and the projectivize constellation space. 
{\sl Generically, distinct shape theories and shape-and-scale theories are not supported}: an argument against ascribing fundamentality specifically to shape theories.  

\m 

\n{\bf Corollary 1} Both of the above relational spaces are moreover just {\sl product spaces}. 
{\sl Thus in the Mechanics case, quotienting by groups of continuous transformations is avoided altogether.} 
Indeed, such finite products of topological spaces preserve initial Hausdorffness, second-countability and local Euclideanness, 
so these constellation spaces inherit $\FrC^d$ and $\FrP^{d - 1}$'s manifoldness. 

\m 

\n{\bf Remark 8} There are moreover plenty of further ways in which finite products of $N$ copies of a given manifold 
are not much harder to treat mathematically than the original manifold itslef.  

\m 
 
\n{\bf Remark 9} Nongeneric $\FrC^d$ -- those that have any $\lFrg \neq id$ for which any nontrivial relational theory occurs -- are of measure zero in the space of all possible 
absolute space models $\FrC^d$.

\subsection{Discussion}

\n{\bf Remark 10} The fork between complicated Stratified Manifold Theory and supporting Sheaf Methods on the one hand -- the symmetric absolute space case -- 
and small defects nullifying presence of any symmetry and hence stratification has become a major aspect of the modern Absolute versus Relational Debate. 
The relational side of this debate sometimes moreover goes under the more contemporary name of `seeking Background-Independent foundations of Physics.

\m 

\n One faces a fork between the following. 

\m 

\n a) Accepting absolute space's symmetry.

\m 

\n b) Eschewing absolute space's symmetry by introducing a small defect.

\m 

\n While a) can confer calculability to some of the simpler physical workings [which b) lacks] 
this absolute space symmetry is however imprinted upon the resulting configuration spaces. 
These strata have moreover been shown to be well capable of driving us outside of the remit of Hausdorff-separated topological spaces 
to the analytically far sparser outlands of merely Kolmogorov-separated topological spaces.  

\m 

\n{\bf Remark 11} This is most familiar in the mathematical intractability of maximal collisions O in the $N$-body problem (for d $\geq 2$). 
This case still moreover possesses two helpful features. 

\m 

\n i) It is localized, so one can still calculate the physics away from it. 

\m 

\n ii) It involves physically-questionable modelling assumptions. 
(Such as that point-particle nonrelativistic mechanics holds to arbitrary precision when all the matter in a universe model 
is contracted to a point.  
One can moreover argue that each of QM, SR, and GR intervene to avoid physical realization of the maximal collision's pathologies. 
For instance, finite-extent of particles that are subsequently not perfectly rigid either, 
wave-particle duality, degeneracy pressure and the formation of black holes could enter the actual physical modelling.)

\m 

\n The situation however worsens considerably even just in Flat Affine or Projective Geometry, 
for which even the shape spaces are afflicted with this, including now physically reasonable configurations such as collinearities.
This being a foundationally significant discussion for both Theoretical Physics and geometrical modelling of Image Analysis and Computer Vision, 
it is certainly worth laying out what further options are afforded.  

\m  

\n{\bf Remark 12} A major connection to the Absolute versus Relational Debate is that 
{\sl reduced spaces set up from symmetric absolute spaces have been shown capable of remembering these symmetries even after the corresponding automorphisms are quotiented out}.  
This remembrance of initially-posited symmetry is in the form of strata occurring in the relational description.
This point is moreover to be added to the {\sl capacity to remember incipient absolute space topology} as is already manifest in e.g.\ 
$N$ points on the circle having a very different relational theory from $N$ points on the line \cite{ACirc}. 
(This example was chosen to exhibit topological effects unconnected to strata, since neither of these distinct relational theories has any strata).  
One can surmise that differential-geometric modelling of Relationalism, by such as `Best Matching' \cite{BB82} 
(which I reformulated as a Lie derivative construct \cite{FileR, ABook} -- not just for GR for which this manifest -- but for any other theory as well, Relational Mechanics theories included) 
is not a freeing from {\sl all} features of the incipient absolute space.
For one's reduced mathematical model can remember not only topological features of the absolute space 
(not so surprising given that the relational freeing manoeuvre is merely differential-geometry level) 
but also of symmetry, through its becoming entangled with topology through strata manifesting themselves in the relational configuration space.
In a nutshell, {\sl for relationalists, considering topology has become a necessity in the quest to free Physics of any imprint of absolute space}.  

\m 

\n Unions of configurations over different absolute space topologies, 
and corresponding `sum over topology' physical constructs \cite{Mis-Top, GH92} (maybe within some class) can formally handle this.   
While in practice these are usually too hard to compute with, 
in some modelling situations they are finite or a systematic sum \cite{Top-Shapes, Forth} and approximations are also possible.  
Our point is that it is necessary to do so if one is in the business of removing {\sl all} absolute-or-background imprints from one's formulation of Physics, 
even in the simpler Mechanics setting.  

\m 

\n{\bf Remark 13} Let us wrap up by giving this Section's alternative answer -- along the lines of a) -- to Research Project 41, 
alongside a one-off aside that one particular family of models offers as a third prong c) to the above fork. 

\m  

\n{\bf Alternative answer} In the Relational Mechanics case, if one works with generic $\bupSigma$ -- approach b) -- 
one encounters no nontrivial generalized symmetry groups, thus no quotienting, thus no strata, and thus no need for Sheaf Methods to replace fibre bundle considerations.  

\m 

\n As an aside -- approach c) -- suppose absolute space is a circle $\mathbb{S}^1$ \cite{ACirc} or its $\mathbb{T}^d$ torus generalization 
(including the physically-plausible case of $\mathbb{T}^3$). 
Then there are no strata in the isometric theory either, by cancellation of the isometry group with part of the constellation space's product structure: 
eq.\ (\ref{Cancel}) and \cite{Forth}.  
This is similar to the situation with $\mathbb{R}$ absolute space (though this again is physically implausible due to its spatial dimension being smaller than 3).  

\vspace{10in}

\section{Conclusion}

\subsection{Overview}

\n As a memorable summary, let us recollect that past builders of cathedrals would introduce some small `countertwist defect' 
to avoid overall symmetry so as to `ward off the devil'. 
Analogously, incorporating a small defect in one's incipient absolute space model wards off the devilry of stratification, 
which would usually elsewise occur in the subsequent reduced configuration spaces, 
and on some occasions be compounded by the onset of analytically-sparse merely Kolmogorov-separated \cite{GT09, KKH16} topological space mathematics. 
But the space and configuration space `cathedrals' thus built could no longer be studied by the exact symmetry methods that Theoretical Physics has hitherto greatly relied upon.  

\m  

\n While in the GR case this defect needs to be a topologically-nontrivial twist, 
in the Kendall-type Relational Theory and subsequent Relational Mechanics a merely-geometrical `bump' suffices.   
Such bumps can moreover be inspired by GR-geometry features such as the Sun or the centre of the galaxy's potential wells.  

\m 

\n Models lacking symmetry are moreover generic; 
we gave a precise meaning to this term from considering generalized Killing equations corresponding to geometrically-significant automorphism groups.
This means that nontrivial Kendall-type Relational Theories are of measure zero on the set of all possible absolute space manifolds, 
a useful observation given recent proliferation of at least partly solvable models in this subject \cite{Kendall, Bhatta, FileR, AMech, PE16, A-Series, ACirc, ASphe, Forth}.  
The trivial theories thereupon are mere product spaces rather than quotients, and thus are free of stratification or other non-manifoldness. 
In the GR case, while the generic reduced configuration space -- Wheeler's superspace -- remains a quotient space by the corresponding automorphisms -- diffeomorphisms -- 
this remains free of strata since generic topological manifolds admit solely the trivial isotropy group. 

\m 

\n So 1960's Geometrodynamics \cite{DeWitt67, DeWitt70, Fischer70} and 1980's Kendall Shape Theory \cite{Kendall84, Kendall} built their `cathedrals' without such a defect, 
thus inviting in the `devilry' of stratification.    
1990's Geometrodynamics \cite{FM96} then contemplated `cathedrals with defects': topological twists on their topological 3-space models.  
Keeping pace, 2010's Generalized Kendall-type Relational Theory has now through the current article considered its own `cathedrals with defects': 
geometrical bumps on their models of carrier space (alias absolute space in the physically-realized context).

\subsection{Research Frontiers}

\n{\bf Research Frontier 1} Extend specific considerations of (stratified manifold, sheaf) pairs \cite{Kreck} 
from Type ii) -- locally compact Hausdorff second-countable -- stratified spaces 
  to Type iii) -- merely Kolmogorov-separated rather than Hausdorff-separated -- counterparts. 

\m 

\n{\bf Research Frontier 2} Of the harder Type iii) strata thus encountered in finite point-or-particle Relational Theory, 
the known examples all occur in the presence of {\it non-resolvably non-compact} geometrical automorphism groups being quotiented out.   
On the other hand, compact automorphism groups and resolvably-noncompact groups give the simplest Type i) strata.
`Resolvable' here refers to passing to the centre of mass frame nullifying the translations, or their compactified equivalents in $\mathbb{T}^d$. 
The aim is then to either find a counterexample or to rest this observation on rigorous theorems.  

\m 

\n{\bf Research Frontier 3: Local-and-Approximate Relational Theory} 
Finite point-or-particle Relational Theory being very different on distinct carrier spaces, and taking but a trivial form generically, 
mounts pressure on reconceiving of such Relational Theory in local and approximate terms. 
Upon finding a situation in which {\sl exact} symmetry methods are not applicable, it is rather natural to contemplate using {\sl approximate} symmetry methods instead...
In such a formulation, small differences in the carrier space geometry would be expected to not cause great differences either in shape configurations thereupon 
or in the topology and geometry of this {\sl Local-and-Approximate Shape Theory}'s own versions of configuration spaces.  
In this new setting, one would expect small shapes on $\mathbb{T}^d$ (relative to the identification lengthscale) to not differ much from those on $\mathbb{R}^d$.  
Similarly, one would expect small shapes on $\mathbb{S}^d$ (now relative to the curvature lengthscale) to not differ much from those on $\mathbb{R}^d$ either.   
Flat charts could well play a role in such a theory, with shapes in overlaps of two such charts being approximately the same in a controlled manner.  
Approximate notions of shape, such as Kendall $\epsilon$-collinearity \cite{Kendall84, Kendall} would be central here, 
rather than exact notions of shape such as perfect collinearity. 
Similarly, tight binaries would be considered in place of binary collisions, and so on.  
One would expect {\sl tangent space} automorphisms (rather than manifold ones) to play a role here.
As would not exact but {\sl approximate} Killing vectors, 
various notions of which have already entered the GR literature 
(see \cite{Matzner, York74, AM78} for the original work and \cite{CKBook, Z-KN, H08, V17} for reviews and \cite{Taubes, AKE, H08} for applications).
These would now need to be generalized to the other notions of geometry's generalized Killing vectors, as covered e.g. in \cite{C79, H08}.  

\mbox{ }

\n{\bf Acknowledgments} I thank Chris Isham and Don Page for discussions about configuration space topology, geometry, quantization and background independence. 
I also thank Jeremy Butterfield and Christopher Small for encouragement. 
I thank Don, Jeremy, Enrique Alvarez, Reza Tavakol and Malcolm MacCallum for support with my career. 
This article is dedicated to my former collaborator Prof. Niall \'{o} Murchadha, on the occasion of his Festschrift, 
who is exceptionally good toward students and junior colleagues.


\end{document}